# Controlling light propagation: *a brief review*

D. P. San-Román-Alerigi

*Photonics Laboratory, King Abdullah University of Science and Technology* (KAUST), *Thuwal 21534, Saudi Arabia*

damian.sanroman@kaust.edu.sa

**Abstract**: *Recent results by Hoffman et al. showed that it is possible to build a Negative Refractive Material (NIM) by interleaving InGaAs/AlInAs using current nanofabrication technologies. Their findings provide the first experimental proof of a negative refractive semiconductor long posited by transformation optics. In this paper an overview on semiconductor refractive index theory and negative refraction is presented, focusing on the results of Hoffman et al. and the future of the subject based on the theory of Transformation Optics and its importance to tailor materials electrodynamics properties..*

## *Introduction*

Tailoring the behavior of the electromagnetic field as it propagates through different media is one of the hot research topics today. This precise control opens the door to hyper-planar lenses, invisibility cloaks, difractionnless propagation, lossless confinement by means of negative refractive index and transformation media (complex alteration of the refractive index of pure and compound metamaterials).

Whilst the idea of shaping the interaction of photons with matters is not new (1; 2), for example Maxwell *Fish Eye* and Lunenburg's Lens are known to theoretically break the diffraction limit (3; 4; 5) by means of shaping the diffractive index within the propagation path; it has been over the last five decades that the possibility of achieving this devices in laboratory came several steps forward, mainly due to the advancement in semiconductors fabrication techniques, and hence metamaterials (6).

## *Refractive Index in Semiconductors*

The refractive index and the band-gap energy of pure and compound semiconductor and metamaterials are two fundamental physical quantities which characterize their optical, and electronic, properties. Hence the possible application (photonic crystals, wave guides, detectors) of semiconductors in optical and optoelectronic devices depends on knowing these characteristic values (7).

The different arrangement of ions and free charges on a crystal gives rise to the energy band-gap. These lattice arrangements given by the different elements combined (binary, ternary, etc.) and doping result in different refractive indexes, thus light propagates differently in every semiconductor.

| Semiconductor Material | Refractive index | $E_g$ band-gap at $300\ K$ |
|---|---|---|
| Zn | 2.008 | 3.400 $eV$ |
| ZnS | 2.368 | 3.680 $eV$ |
| ZnTe | 2.720 | 2.394 $eV$ |
| CdS | 2.506 | 2.500 $eV$ |
| CdSe | 2.500 | 1.714 $eV$ |

Table 1. Semiconductors and their refractive index with different dopants **(8)**.

Recall that the refractive index is related to the dielectric and permittivity constants by the equation:

$$n = \pm\sqrt{\epsilon\mu}$$

Eq. 1

For most materials and semiconductors the permittivity $\mu$ is close to one, so the diffractive index is in most cases dependent on the dielectric constant of the material, which in turns depends on the free and bound electrons of it. Accounting for the free electrons in an isotropic semiconductor it can be shown that the dielectric function is (7):

$$\epsilon \approx 1 + \left(\frac{\hbar\omega_p}{E_0}\right)^2 + \left[1 - \frac{E_g}{4E_F} + \left(\frac{E_g}{E_F}\right)^2\right] + 48$$

Eq. 2

Where $\omega_p$ is the plasma frequency (resonance frequency of electrons) and $E_F$ the Fermi energy. This method however can only account for standing waves near the Brillouin zone, yet since the model is based on the isotropy of the system it can be used to study amorphous semiconductors.

As shown in Table 1 the band-gap energy and the refractive index are entwined; the first to posit the relation between these two quantities was Moss in 1950 (9) who proposed the following relation for the refractive index $n$ and the the energy band gap $E_g$

$$n^4 E_g = constant$$

Eq. 3

The constant in Eq. 3 has had different values (95, 108 and 173) (10).

Following Moss and Penn relations in 1979 and 1980 Ravindra, Gupta *et al.* proposed a new relation for the refractive index and the band-gap energy based initially on empirical research (11; 12; 13):

$$n = 4.16 - 1.12 E_g + 0.31 E_g^2 - 0.08 E_g^3$$

Eq. 4

| Semiconductor | Real Value | Moss Model | Ravindra, Gupta *et al.* model |
|---|---|---|---|
| PbSe | 4.7 | 4.6 | 0 |
| PbSnTe | 7.0 | 6.03 | 0 |
| GaN | 2.29 | 2.41 | 2.1 |

Table 2. Comparison of Moss and Ravindra-Gupta model for refractive index (7).

Ravindra and Gupta based their study on a common feature to all semiconductor band structures in which the valence and conduction bands are almost parallel along the symmetry directions. A very important drawback of this relation is that predicts a zero refractive index value beyond $n \geq 4.1$, where some important infrared materials are found.

Based on Moss and Ravindra's relation Herve and Vandamme proposed a model to account for the fact that some semiconductors are ionically bonded and others are covalently bonded. For the later they found that the dielectric function is:

$$\epsilon(\omega) = 1 + \frac{4\pi N q^2}{m(\omega_0^2 - \omega^2)}$$

Eq. 5

Where $\omega_0$ is the UV resonance frequency, $m$ is the mass at rest of the electrons and $N$ is the density of electrons. These equations fit the experimental results for Ge and Si and for almost all the covalently bounded materials. For ionically bonded, like GaAs, the dielectric function is dependent on the plasma frequency of ions $\Omega_p$ and the infrared resonance frequency $\omega_T$, taking into account the optical range $\omega_{op}$ ($\omega_T \leq \omega_{op} \leq \omega_0$) the dielectric function takes the form:

$$\epsilon(\omega) = \epsilon(\omega_{op}) + \frac{\Omega_p^2}{\omega_T^2 - \omega^2}$$

Eq. 6

Finally Herve and Vandamme proposed for the optical regime a simplified equation:

$$n = \sqrt{1 + \frac{13.6^2}{(E_g + B)^2}}$$

Eq. 7

Where $B$ has the value $3.47\ eV$ except for resonances in the UV where it takes the value:

$$\hbar \omega_0 = E_g + B$$

Eq. 8

This last model provides the lowest deviation for III-V and I-VII semiconductors (10; 14), however is unable to provide a correct prediction for IV-VI semiconductors (7).

The Herve-Vandamme model is accurate when compared to the experimental results, however they later showed (15; 7) that Moss relation is accurate for band-gap energies greater than $1.43\ eV$; and that Vendra-Gupta relation provide the best fit for materials with band-gap energy smaller than $1.43\ eV$ (13).

Whilst the former models yield an accurate result for simple materials as the materials grow in complexity different methods, numerical and experimental, are used to calculate its electronic and optical properties. For example programs like Crystal and Espresso can be used to derive the dielectric tensors of different compounds. As well experimental methods like Z scan and spectroscopy are used to derive the optical constants for different materials.

*Negative refractive index*

A negative refraction material (NIM) or left handed material (LHM) is one where the permittivity $\epsilon$ and the permeability $\mu$ are simultaneously negative, observe that the refractive index, Eq. 1, will still be real.

Negative refractive index was first proposed by Veselago in 1968 (16). Yet it was until 2000 that a metamaterial fabricated by Smith *et al.* (17) exhibited negative refraction.

It can be shown (16; 18) that whereas in a positive refractive index material (PIM) the group ($v_g$) and phase ($v_{ph}$) velocities coincide with the direction of the Poynting vector $S$ and the wave vector $k$, in a NIM the Poynting vector and the wave vector are antiparallel, moreover the phase velocity coincides with the direction of the wave vector and the group velocity with that of the Poynting vector.

The former behavior has a striking consequence on Snell Law, for positive $\epsilon$ and $\mu$ the ray propagates following the path 1-3, whereas for negative $\epsilon$ and $\mu$ it takes the path 1-4 (19; 18). The former arises from the fact that the group and phase velocities are antiparallel, and due to the continuity (boundary conditions) of the tangential components of the electric field along the interface between the two media.

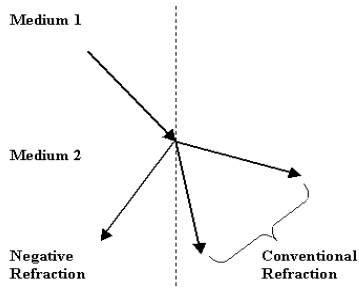

Fig. 1. Graphical description of negative and conventional refractive index materials.

Recalling Snell Law the former result implies that the negative sign in **Eq. 1** should be taken:

$$n_1 \sin\theta_1 = -n_2 \sin\theta_2$$

Where $\theta_1$ is the incident angle and $\theta_2$ is the angle of the refractive ray.

A more detailed explanation and the consequences of negative refractive index materials are beyond the scope of this review, the interested reader can refer to (18; 19)

*NIM Fabrication and semiconductors*

Many of the recent attempts to create NIMs rely on overlapping electronic and magnetic resonances such that $\epsilon, \mu < 0$ CITATION RAS \l 1033 \m TKo041| (20; 21)}. However the design and fabrication of such metamaterials is highly complex and the possibility to implement them in a complete 3D configuration is limited by fabrication technology.

We have studied that the refractive index of semiconductors depends on the materials (III-V, IV-VI) and the doping, i.e. the band gap; moreover current fabrication technology allows for $90\ nm$ resolution, this high sophistication in manufacture and our understanding of semiconductor materials either seldom or in more complex structures (transformation media) is capital to overcoming the hitherto dificulties in fabrication of materials with negative refraction. Hoffman et al. (22) used these knowledge to demonstrate a semiconductor metamaterial with negative refraction for all angles of incidence in the whole long-infrared ($> 1\ \mu m$) spectrum. The material posited consists on alternating layers of $InGaAs$ and intrinsic $AlInAs$. This is an anistropic material which exhibits an interesting property: the starting wavelenght is controlled by the electron density in the doped layers (22).

As shown in Fig. 2 $InGaAs/AlInAs$ exhibits anisotropy, i.e. it has different permitivity, hence refractive index, along different crystal directions.

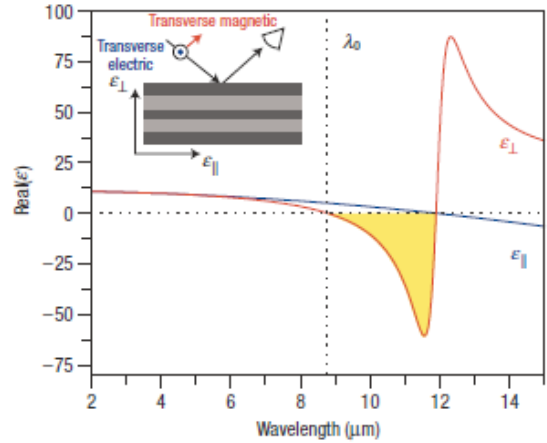

Fig. 2. Schematic design of the AlGaAs/AlInAs NIM and the permittivity constants.

In their experiment Hoffman et al. used a material composed of interleaved $In_{0.53}Ga_{0.47}As$ $80\ nm$ thick and $Al_{0.48}In_{0.52}As$, grown by MOCVD on latticed matched $InP$. Anisotropy is achieved by injection carriers to dope the $InGaAs$ layers, notice that free carriers will result in a plasma resonance frequency, very much like ideal metals (23; 24; 6). This resonance results in negative refraction for certain wavelengths (yellow shade in Fig. 2).

Anisotropy is key to this configuration, recall that in such medium $E$ and $S$ are not necessarily parallel, then the wave vector, $k$, and the Poyinting vector can point in different directions. For an uniaxial medium such as this semiconductor $\epsilon_\parallel > 0$ but $\epsilon_\perp < 0$, hence $k$ and $S$ are antiparallel, i.e. whilst the wave effectively moves in the positive direction the energy flux, Poyinting vector, is in the negative direction (25; 19).

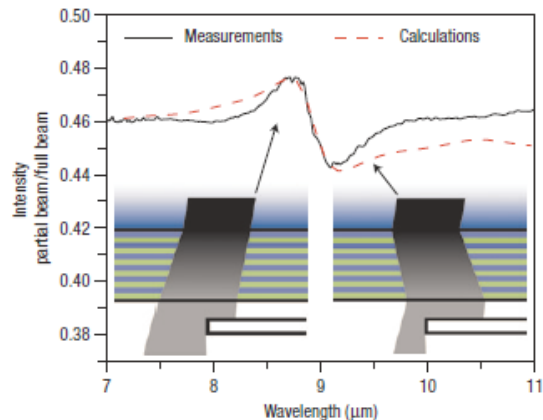

Fig. 3. Schematic representation and results for transmission intensity of the full beam (TM, TE) and for partial beam (TM) through the material.

For this kind of material only the transversal component is negative, Fig 2, i.e. the $TM$ mode. As we have discussed doping plays an important role in negative dielectric constant for the given wavelength range, however highly doped semiconductors has as a side effect: absorption; in this experiment the absorption $\alpha_{TM}$ for doping $n_d = 7.5 \times 10^{-19} cm^{-3}$ is between $1800\ cm^{-1}$ to $2100\ cm^{-1}$ as function of the incidence angle (22).

By engineering the dielectric function (doping, stacking/interleaving different semiconductors) of anisotropic semiconductor materials is then possible to achieve NIMs across wide wavelength range (25)., albeit current technologies allow mainly in the infrared regime. Yet this approach being based on planar semiconductor and based on current growth techniques overcomes previous difficulties opening the door for low-loss (comparatively to previous methods) three dimensional negative refractive index materials which will allow for new application in telecommunications and imaging applications; for example breaking the diffraction limit, which will result in perfect focusing, smaller ($10\ nm$) size lithography, increased resolution for imaging, telecommunication optical switches.

Hoffman *et al.* are not the only group who showed this for semiconductors and metamaterials, following their discovery in 2010 a model accounting for quantum photon-matter interaction proposed metamaterial with negative refractive index (25), implemented on thin films of tin-doped indium oxide (ITO) .

*Transformation Media – Transformation Optics*

The theoretical background to study effects such as NIM in complex materials relies on Maxwell electrodynamics theory and within to the special topic of Transformation Optics or Transformation Media, to which Pendry and Leonhardt have significantly contributed over the last decade (26)

Transformation optics relies on the fact that negative refraction, cloaking and super lenses are examples where materials seem to transform the geometry of the space and thus shape the way light propagates.

A brief mathematical description is as follows: assume that light propagates in empty space $\mathbf{A} = (x, y, z)$ which has a certain refractive index $n_R$. As the light travels it may interact with any variety of objects that will result in diffraction, refraction and dispersion; all these can be seen as mappings of the waves by some function $f$ to a space $\mathbf{B} = (x', y', z')$, with refractive index $n_r(x', y', z')$, i.e. permittivity tensor $\epsilon'^{i'j'}$ and permeability tensor $\mu'^{i'j'}$, such that $\mathbf{B} = f(\mathbf{A})$, observe that since both spaces are different this implies that in general the geometric metric of both spaces is different, i.e. $dx^2 + dy^2 + dz^2 \neq dx'^2 + dy'^2 + dz'^2$. The metric depends on the characteristic of the medium, i.e. on the dielectric tensor $\epsilon$ and the permeability $\mu$, hence on the refractive index and dispersion characteristics of it (27).

Is important to note that the transformation $f$ is not arbitrary since it should comply with all the basic physical laws, this means that some desired effects while mathematically possible may not be physically achievable.

In general any transformation from a set of coordinates to another metric can be described using Einstein notation: $x'^{i'j'} = f(x^{ij})$. The transformation $f$ is then a change of coordinates characterized by the Jacobian which elements are:

$$\Lambda_j^{j'} = \frac{\partial x'^{j'}}{\partial x^j}$$

Eq. 9

Thus the permeability and permittivity change to satisfy Maxwell Equations:

$$\epsilon'^{i'j'} = [det(\Lambda)]^{-1} \Lambda_i^{i'} \Lambda_j^{j'} \epsilon^{ij}$$
$$\mu'^{i'j'} = [det(\Lambda)]^{-1} \Lambda_i^{i'} \Lambda_j^{j'} \mu^{ij}$$

Eq. 10

This new set of permittivity and permeability describe the propagation of light in the given medium. Thus if one desires to have a negative refractive medium is necessary to find the accurate transformation, i.e. the new $\epsilon$ and $\mu$ (26; 27). We know that that the transformation for NIM is one where Snell Law is left handed, thus the transformation $f$ is a rotation which magnitude depends on the angle desired.

The former means that if one knows the final and initial states then the problem reduces to find the transformation which yields this result and with it finding $\epsilon$ and $\mu$. This transformation $f$ chosen properly will result in invisibility or perfect focusing to a single point or negative refraction.

Now let us think about an invisibility cloak, to make two metrics the same the mapped space $\mathbf{B}$ should be an optical medium such that both metrics are equal. This means that the object we want to cloak should be embedded or surrounded by it. Hence light will follow the images $f(l)$ of the straight lines $l$ of the original propagation space.

Based on Transformation optics many groups had proposed different devices made of semiconductor and metamaterial compounds, for example Yun Lai *et al*. (28) theoretically showed that is possible to cloak large objects and even make cloak at a distance.

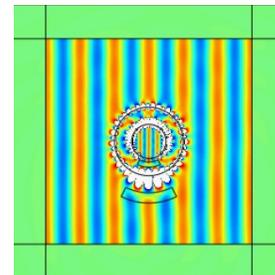

Fig. 4. Simulation of an Invisibility cloak using transformation media.

The device shown in Fig. 4 Is currently under study to take it to the laboratory, it can be based on an interleaved gold and highly doped semiconductor material (Si for example) .

Another interesting device demonstrated by transformation media could reproduce the diffraction of an object, thus making possible to change the diffraction pattern of a an object into another object (29).

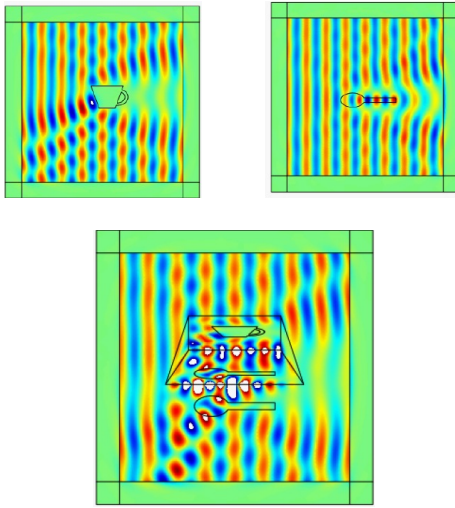

Fig. 5. Diffraction patter of a spoon and a cup (Top). Diffraction patter of spoon changed to the cup by means of complimentary media (Bottom)

The former device is highly theoretical, however it shows that tailoring the dielectric tensor could result in *illusion optic* device. This device could be achieved by interleaving different semiconductor and metamaterials to achieve the anisotropy required.

*Conclusions*

As noted by Shalaev "Transformation optics, which includes NIMs, is a new way of manipulation and controlling light at all distances, from the macro to the nanosclae, and it represents a new paradigm for the science of light".

With the aid of transformation optics one can render an object invisible by curving the light around it or concentrate all the light in a point which will result in a powerful tool for telecommunication, imaging and solar energy applications

Negative refractive materials and the possibility to fabricate them as shown by Hoffman *et al.* using semiconductor materials and fabrication technology is an important breakthrough for a research topic that until very recently it had rely in theoretical models and simulations. More research, however, is needed to improve their performance in the optical regime where these devices have many applications, mainly in imaging and optical telecommunication.

We have also analyzed Transformation Optics, noting that it provides a theoretical tool to analyze different electrodynamics properties and propagation of light in complex metamaterials and semiconductors. The theory allows finding the necessary permittivity and permeability tensors a material should have to achieve different devices.

Observe that whilst Transformation Optics is mathematically elegant the difficulty arises in finding materials which can have the specified electromagnetic tensors, hence the importance of Hoffman *et al.* experiment. Their research proved that already accessible techniques for semiconductor fabrication and doping can be used to engineer the desire characteristics, still research is need to improve this fabrication in order to allow for design in the tens of nanometer regime to produce the same results in the whole optical spectrum.

**Works Cited**
1. *Invisible Bodies.* **Kerker, M.** s.l. : J. Opt. Soc. Am., 1975, Vol. 65, p. 376.
2. *Anbornally low electromagnetic scattering cross sections.* **H. Chew, M. Kerker.** s.l. : J. Opt. Soc. Am., 1976, Vol. 66, p. 445.
3. **Lenz, W.** Probleme der Modernen Physik. Leipzig, Hirzel : P. Debye, 1928, p. 198.
4. **Maxwell, J. C.** s.l. : Cambridge and Dublin Math J., 1854, Vol. 8, p. 188.
5. **M. Born, E. Wolf.** *Principles of Optics.* s.l. : Cambridge University Press, 2006.
6. **S. Zouhdi, A. Sihvola, A. P. Vinogradov.** *Metamaterials and Plasmonics: Fundamentals, Modeling, Applications.* New York : Springer-Verlag, 2008.
7. *Energy gap-refractive index relations in semiconductors - An overview.* **N. M. Ravindra, P. Ganapathy, J. Choi.** s.l. : Elsevier. Inf. Phys & Tech., 2006, Vol. 50, p. 21.
8. **Palmer, D. W.** *The Semiconductors-Information Web-Site.* [Online] University of Exeter, January 2009. http://www.semiconductors.co.uk/.
9. **Moss, T. S.** s.l. : Proc. Phys. Soc., 1950, Vol. B63, p. 167.
10. *Energy gap-refractive index interrelation.* **Gopal, V.** s.l. : Elsevier. Infrared Phys., 1982, Vol. 22, p. 255.
11. *Variation of refractive index with energy gap in semiconductors.* **N. M. Ravindra, V. K. Srivastava.** s.l. : Elsevier. Infrared Phys., 1979, Vol. 19, p. 603.
12. **N. M. Ravindra, S. Auluck, V. K. Srivastava.** s.l. : Wiley. Phys. Stat. Sol., 1979, Vol. 93, p. 155.
13. *Comments on the Moss formula.* **V. P. Gupta, N. M. Ravindra.** s.l. : Wiley. Phys. Sol., 1980, Vol. 100, p. 715.
14. *Correlation between the refractive index and the energy gap of simple and complex binary compounds.* **R. P. Singh, P. Singh, K. K. Sarkar.** s.l. : Elsevier. Infrared Phys., 1986, Vol. 26, p. 1.
15. **P. J. L. Herve, L. K. J. Vandamme.** s.l. : J. App. Phys., 1995, Vol. 77.
16. *Electrodynamics of substances with simultaneously negative values of sigma and mu.* **Veselago, V.G.** s.l. : Sov. Phys. Usp., 1968, Vol. 10, p. 509.
17. **D. R. Smith, W. J. Padilla, D. C. Vier, S. C. Nemat-Nasser, S. Schultz.** 2000, Phys. Rev. Lett., Vol. 84, p. 4184.
18. *Negative refractive index materials.* **V. Veselago, L. Braginsky, V. Shklover, C. Hafner.** s.l. : J. Comp. Theoretical Nanoscience, 2006, Vol. 31, p. 1.
19. *Physics of negative refractive index materials.* **Ramakrishna, S. A.** s.l. : Rep. Prog. Phys., 2005, Vol. 68, p. 449.
20. *Experimental verification of negative index refraction.* **R. A. Shelby, D. R. Smith, S. Schultz.** Science, Vol. 292, p. 77.
21. *Effective medium theory of left handed materials.* **T. Koshny, M. Kafesaki, E. Economou, C. M. Soukoulis.** 2004, Phys. Rev. Lett., Vol. 93, p. 107402.
22. *Negative refraction in semiconductor metamaterials.* **A. J. Hoffman, L. Alekseyev, S. S. Howard, K. J. Franz, D. Wasserman, V. A. Podolskiy, E. E. Narimanov, D. L. Sivco, C. Gmachl.** s.l. : Nature, 2007, Vol. 10.
23. *Injected-carrier induced refractive index change in semiconductor.* **A. Olsson, C. L. Tang.** s.l. : App. Phys. Lett., 1981, Vol. 39, p. 24.
24. **N. Engheta, R. W. Ziolkowski.** *Metamaterials: physics and engineeringexplorations.* s.l. : Wiley & Sons, 2006.
25. *Negative refractive index in doped semiconductors.* **A. G. Kussow, A. Akyurtlu.** s.l. : Proc. App. Phys. Soc. March Meeting 2010, 2010. BAPS.2010.MAR.B14.9.
26. *Focus on cloaking and transformation optics.* **U. Leonhardth, D. R. Smith.** s.l. : New J. Phys., 2008, Vol. 10, p. 115019.
27. **U. Leonhardth, T. G. Philbin.** s.l. : New. J. Phys., 2006, Vol. 8, p. 247.
28. *Complementary media invisibility cloak that cloacks objects at a distance ouside the cloaking shell.* **Y. Lai, H. Chen, Z. Zhang, C. T. Chan.** s.l. : Phys. Rev. Lett., 2009, Vol. 102, p. 093901.
29. *Illusion optics: the optical transformation of an object into another object.* **Y. Lai, J. Ng, H. Chen, D. Han, J. Xiao, Z. Zhang, C. T. Chan.** s.l. : Phys. Rev. Lett., 2009, Vol. 102, p. 253902.

30. *Composite medium with simultaneously negative permeability and permitivity.* **D. R. Smith, W. J. Padilla, D. C. Vier, S. C. Nemat-Nasser, S. Schultz.** s.l. : Phys. Rev. Lett., 2000, Vol. 84, p. 4184.
31. **G. V. Eleftheriades, K. G. Balmain.** *Negative-Refraction Metamaterials.* s.l. : Wiley-IEEE Press, 2005. p. 440.
32. **Moss, T. S.** Wiley. Phys. Stat. Sol. : s.n., 1985, Vol. 131, p. 415.
33. **P. J. L. Herve, L. K. J. Vandamme.** s.l. : Wiley. Infrared Phys., 1994, Vol. 35, p. 609.
34. *Optical Conformal Mapping.* **U. Leonhardth, et al.** s.l. : Science, 2006, Vol. 312, p. 1777.
35. *Broadband invisibility by non-euclidian cloaking.* **U. Leonhardth, T. Tyc.** s.l. : Science, 2009, Vol. 323, p. 110.
36. *Metamatirals in electromagnetics.* **Sihvola, A.** s.l. : Elsevier, 2007, p. 2. DOI:10.1016/j.metmat.2007.02.003.